# A Script Language for Data Integration in Database

Qingguo Zheng

## Abstract

   A Script Language in this paper is designed to transform the original data into the target data by the computing formula. The Script Language can be translated into the corresponding SQL Language, and the computation is finally implemented by the first type of dynamic SQL. The Script Language has the operations of insert, update, delete, union, intersect, and minus for the table in the database.
    A Script Language is edited by a text file and you can easily modify the computing formula in the text file to deal with the situations when the computing formula have been changed. So you only need modify the text of the script language, but needn't change the programs that have complied.

## The format of Script Language

```
{
 TABLE: table_name
 COMMAND:command_name
 Target_item_one=[DISTINCT]Source_item||function(…)[<operator>Source_item |function(…)]
       …          ==         ….
 Target_item==Source_item|function(…)[<operator>Source_item |function(…)]
 Target_item_n==Source_item| function(…) [<operator> Source_item |function(…)]
 [FROM:<Source_table>[,<Source_table>]]
 [WHERE:<Condition>[,<condition>]]
 [GROUP BY: <group_item>[，<group_item>]]
 [HAVING]:<Condition>[,<Condition>]]
 [ORDER BY:<order_item>[,<order_item>]]
}
[set_operator]
```

```
{
 TABLE: table_name
 COMMAND:command_name
 Target_item_one==[DISTINCT]Source_item||function(…)[<operator>Source_item |function(…)]
       …        ==      ….
 Target_item==Source_item|function(…)[<operator>Source_item |function(…)]
 Target_item_n==Source_item| function(…) [<operator> Source_item |function(…)]
 [FROM:<Source_table>[,<Source_table>]]
 [WHERE:<Condition>[,<condition>]]
 [GROUP BY: <group_item>[，<group_item>]]
 [HAVING]:<Condition>[,<Condition>]]
 [ORDER BY:<order_item>[,<order_item>]]
}
set_operator=[UNION,INTERSECT,MINUS]
operator=[<+>,<-,><*>,</>]
command_name=[insert,update,delete,truncate,create]
说明：
         function(…): the parameter are the tuples of the source table.
       '{' represents the begin of one operation.
       '}' represents the end of one operation.
       '//' represents the next line is the same line with current line.
       '#' represents after '#' notion is the explaining description
       '/*' and '*/' represents the text from '/*' to '*/' is the explaining description.
```

## The represent of the repeated part in UNION operation

```
{
 TABLE: table_name
 COMMAND:command_name
 Target_item_one=[DISTINCT]Source_item||function(…)[<operator>Source_item |function(…)]
       …        ==      ….
 Target_item==Source_item|function(…)[<operator>Source_item |function(…)]
 Target_item_n==Source_item| function(…) [<operator> Source_item |function(…)]
 [FROM:<Source_table>[,<Source_table>]]
 [WHERE:<Condition>[,<condition>]]
 [GROUP BY: <group_item>[，<group_item>]]
 [HAVING]:<Condition>[,<Condition>]]
 [ORDER BY:<order_item>[,<order_item>]]
}
UNION:
{
TABLE: table_name
 COMMAND:command_name
     SAME:==SAME:
```

[FROM:<Source_table>[,<Source_table>]]
  [WHERE:<Condition>[,<condition>]]
  [GROUP BY: <group_item>[，<group_item>]]
  [HAVING]:<Condition>[,<Condition>]]
  [ORDER BY:<order_item>[,<order_item>]]
}

## the format of UPDATE operation

{
 TABLE: table_name
 COMMAND:UPDATE
 Target_item_one==value_item||function(…)[<operator>value_item |function(…)]
         …          ==          ….
 Target_item==value_item|function(…)[<operator>value_item |function(…)]
 Target_item_n==value_item| function(…) [<operator> value_item |function(…)]
 [WHERE:<Condition>[,<condition>]]
}
Corresponding SQL：
            update    table_name
              set
              Target_item_one=value_item||function(…)[<operator>value_item |function(…)]
       …          =          ….
              Target_item=value_item|function(…)[<operator>value_item |function(…)]
              Target_item_n=value_item| function(…) [<operator> value_item |function(…)]
              WHERE:<Condition>[,<condition>]

## The nested format of UPDATE operation

{
 TABLE: table_name
 COMMAND:UPDATE
{
 Target_item_1==[DISTINCT]Source_item||function(…)[<operator>Source_item |function(…)]
         …          ==          ….
 Target_item_2==Source_item|function(…)[<operator>Source_item |function(…)]
 Target_item_n==Source_item| function(…) [<operator> Source_item |function(…)]
 [FROM:<Source_table>[,<Source_table>]]
 [WHERE:<Condition>[,<condition>]]
}
   [WHERE:<Condition>[,<condition>]]

}
Corresponding SQL:

                        update    table_name
                        set (Target_item_1, Target_item_2,…, Target_item_n)=(select
                         ]Source_item||function(…)[<operator>Source_item ],…,
                         Source_item||function(…)[<operator>Source_item ]
                         FROM:<Source_table>[,<Source_table>]
                         WHERE:<Condition>[,<condition>]    )
               WHERE:<Condition>[,<condition>]

## The format of creating views and tables

- Creating Views

{
TABLE: table_name
 COMMAND:CREATE VIEW
 Target_item_one==[DISTINCT]Source_item||function(…)[<operator>Source_item |function(…)]
         …         ==         ….
 Target_item==Source_item|function(…)[<operator>Source_item |function(…)]
 Target_item_n==Source_item| function(…) [<operator> Source_item |function(…)]
 [FROM:<Source_table>[,<Source_table>]]
 [WHERE:<Condition>[,<condition>]]
 [GROUP BY: <group_item>[，<group_item>]]
 [HAVING]:<Condition>[,<Condition>]]
 [ORDER BY:<order_item>[,<order_item>]]
}
Corresponding SQL: create view   table_name   as
                         select   Source_item|[Formula]，…, Source_item|[Formula]
                         FROM:<Source_table>[,<Source_table>]
                         WHERE:<Condition>[,<condition>]

- Creating table
1. Create the basic table

{
TABLE: table_name
 COMMAND: ONLY CREATE
 Target_item_1==[NUMBER …[VARCHAR]]
      …       ==       ….
 Target_item_2==[NUMBER …[VARCHAR]]
 Target_item_n==[NUMBER …[VARCHAR]]
}

Corresponding SQL: create table  table_name(target_item_1   [NUMBER …[VARCHAR]] ,…
                                , Target_item_n   [NUMBER …[VARCHAR]]   )
2.  Creating table by replicating other table

{
TABLE: table_name
 COMMAND:CREATE
 Target_item_one==Source_item|
      …        ==        ….
 Target_item==Source_item
 Target_item_n==Source_item
 [FROM:<Source_table>[,<Source_table>]]
 [WHERE:<Condition>[,<condition>]]
 }
Corresponding SQL: create table table_name as select   Source_item|[Formula],
                        …, Source_item|[Formula]
                        FROM:<Source_table>[,<Source_table>]
                        WHERE:<Condition>[,<condition>]

## The format of deleting data

- delete operation
{
TABLE: table_name
COMMAND:delete
[WHERE:<Condition>[,<condition>]]
}
Corresponding SQL: delete table_name where   condition
- truncate operaion
{
TABLE: table_name
COMMAND: truncate
}
Corresponding SQL: truncate   table   table_name;

# The operation of Dropping tables and views

- Dropping tables
{
TABLE: table_name
COMMAND: drop
}
Corresponding SQL：drop table table_name;
- Dropping views
{
TABLE: table_name
COMMAND: drop view
}
Corresponding SQL：drop view table_name;

# Creating the index of table

{
TABLE:index_name
COMMAND:create index
Table_name==attribute_1_name
Table_name==attribute_2_name
 …. ==…
Table_name==attribute_3_name
}
Corresponding SQL：

Create index index_name
on table_name(attribute_1_name,attribute_2_name,…,attribute_3_name)

# Inserting the values into the table

{
TABLE: table_name
COMMAND:insert
Target_item_1==values_1
 …       ==     ….
Target_item_2==values_2
Target_item_n==values_n
}
Corresponding SQL：insert into table_name(Target_item_1,…, Target_item_n)
values(values_1,…, values_n)

# When Target_item is '*', representing all the tuples in the target table

**Example 1.**
```
{
  table:drop_call
  command:insert
   *=distinct item1
   *=item2
   *=item3
   from:table_name
   where: item1>80
}
```
Corresponding SQL: insert into table_name select distinct item1,item2,item3 from table_name where item1>80

**Example 2.**
```
{
  table:drop_call
  command:insert
   *=*
   from:table_name
   where: item1>80
}
```
Corresponding SQL:
insert into table_name select * from table_namewhere item1>80

**Example 3.**
```
{
  table:drop_call
  command:insert
  target_item1==percent(source_item_1,source_item_2)
  target_item2==source_item_3+source_item_4
  target_item3==source_item_6
  from: table_n_1
  where source_item>89
}
```
Corresponding SQL:
insert into drop_call(target_item1,target_item2,target_item3) select percent(source_item_1,source_item_2),source_item_3+source_item_4,source_item_6 from table_n_1 where source_item_5>89
where percent(item1,item2) is function of computing item1 divided by item2.

**Example 4.**
```
{
  table:drop_call
  command:delete
  from: table_n_1
  where source_item>89
}
```
Corresponding SQL：
   delete    from table_n_1    where source_item_5>89

**Example 5.**
```
#1* select text from user_views where view_name='V_P_T_GOS'
{
table:gos_test
command:insert
NETWORK==0 NETWORK
OVERFLOW==A.TCH_REQ_REJ_LACK-A.TCH_REJ_UND_OVER  OVERFLOW
TOTAL==A.TCH_REQUEST-A.TCH_REQUEST_UND_OVER TOTAL
SHIJIAN==TO_CHAR(A.PERIOD_START_TIME,'YYYYMMDDHH24') SHIJIAN
BTS_NAME==B.NAME BTS_NAME
BTS_ID== B.INT_ID BTS_ID
FROM:\\
  OMC.TEMP_TRAFFIC A,\\
  NMC.CENTRAL_OBJECT_INFO_TABLE B
WHERE:\\
  B.OMC_ID=1 AND
  A.BTS_INT_ID=B.NOKIA_ID
}
UNION
{
table:gos_test
command:insert
NETWORK==0 NETWORK
OVERFLOW== A.TCH_REQ_REJ_LACK-A.TCH_REJ_UND_OVER  OVERFLOW
TOTAL==A.TCH_REQUEST-A.TCH_REQUEST_UND_OVER TOTAL
SHIJIAN==TO_CHAR(A.PERIOD_START_TIME,'YYYYMMDDHH24') SHIJIAN
BTS_NAME== B.NAME BTS_NAME
BTS_ID== B.INT_ID BTS_ID
FROM:\\
  OMC1.TEMP_TRAFFIC A,\\
  NMC.CENTRAL_OBJECT_INFO_TABLE B
WHERE:\\
  B.OMC_ID=2 AND\\
  A.BTS_INT_ID=B.NOKIA_ID
```

}